\documentclass[reprint, amsmath,amssymb, aps, showpacs, pra]{revtex4-1}

\usepackage{graphicx}
\usepackage{dcolumn}
\usepackage{bm}

\def\bra#1{{\langle #1 |}}
\def\ket#1{{| #1 \rangle}}

\begin{document}
\preprint{}

\title{Low error measurement-free phase gates for qubus computation}

\author{T. J. Proctor}
 \email{py08tjp@leeds.ac.uk}
\author{T. P. Spiller}
 \email{t.p.spiller@leeds.ac.uk}
\affiliation{School of Physics and Astronomy, E C Stoner Building, \\ University of Leeds, Leeds, LS2 9JT, UK}

\date{\today}

\begin{abstract}
We discuss the desired criteria for a two-qubit phase gate and present a method for realising such a gate for quantum computation that is measurement-free and low error. The gate is implemented between qubits via an intermediate bus mode. We take a coherent state as the bus and use cross-Kerr type interactions between the bus and the qubits. This new method is robust against parameter variations and is thus low error. It fundamentally improves on previous methods due its deterministic nature and the lack of approximations used in the geometry of the phase rotations. This interaction is applicable both to solid state and photonic qubit systems. 

\end{abstract}

\pacs{03.67.Lx, 03.67.-a, 32.80.-t, 42.50.-p}

\maketitle

\section{Introduction}
The \emph{qubus} computational architecture has considerable promise for helping to overcome some of the problems associated with developing a scalable quantum computer. The qubus computer employs a bus mode to mediate the interactions between computational qubits, removing the need for direct qubit-qubit interactions \cite{Milburn06}. In the case of solid state qubits these direct interactions often require close proximity between the qubits due to the nature of the interactions used. This close proximity can make applying individual control fields and measuring the state of individual qubits extremely challenging. Furthermore creating entanglement or enacting gates between non-adjacent qubits can create large SWAP gate overheads when only nearest-neighbour interactions are available. The use of a bus to mediate the interactions can remove these problems. It is clear therefore that the qubus architecture has the potential to improve the viability of solid state computation. However, a wider advantage of this approach is that it isn't necessarily restricted to a particular type of computational qubit. A qubus can be used to mediate between solid state qubits, creating the entanglement needed for measurement based quantum computation, but equally it can be employed with photonic qubits and the original gate model of quantum computation. It could even potentially be used to mediate between different types of quantum architecture. 
\newline  \indent Here we will consider using a coherent state of light as the bus mode. This has the advantage of employing a continuous quantum variable for communication \cite{Braunstein05} \cite{Lloyd99}, and this could even potentially be used to communicate with existing classical technologies. The aim therefore is to develop methods for implementing the entangling operations or multi-qubit gates required for quantum computation using a bus mode. Clearly a good starting point is the development of a controlled phase (C-Phase) gate as this, in conjunction with single qubit rotations, is a universal set of gates for quantum computation \cite{Ekert95}. 
\newline  \indent To achieve the required bus-qubit interactions we will first consider photonic qubits and the cross-Kerr optical non-linearity between two modes $a$ and $b$. This has a Hamiltonian of the form
\begin{equation}
\label{Hint}
H_{int}=\hbar \chi a^{\dagger}a b^{\dagger} b
\end{equation}
where $a$ ($a^{\dagger}$) refers to the annihilation (creation) operator of the bus electromagnetic field mode, $b$ ($b^{\dagger}$) corresponds to the annihilation (creation) operator of a photonic qubit mode and $\chi$ is the coupling strength. The interaction $H_{int}$ applied for a time $t$ generates either a phase rotation of $\theta \equiv \chi t$ or no phase rotation on the bus field mode dependent upon the qubit being in the number ($b^{\dagger}b$) eigenstate, $\ket{1}$ or $\ket{0}$ respectively.  When the non-linearity is large ($\theta$ is of order $\pi$) this Hamiltonian naturally implements a C-Phase gate. This gave rise to many proposals for its application, such as N00N-state generation \cite{Gerry01} and optical universal quantum computers \cite{d'Ariano00}. However naturally occurring Kerr media have a dimensionless interaction magnitude of order $\theta \approx 10^{-18}$ for realistic interaction times \cite{Kok02}. Despite this, it is possible to fabricate materials with $\theta \approx 10^{-2}$ using techniques such as Electromagnetically Induced Transparencies (EIT) \cite{Yang09} \cite{Sun08} \cite{Spiller05} \cite{Harris99}, optical fibers \cite{Matsuda09} \cite{Li04} and cavity QED systems \cite{Zhu10} \cite{Mucke10} \cite{Dao-Ming07} and there is very significant research and recent progress in many of these areas. However, it should also be noted that phase noise and photon loss need careful 
consideration and may form potential issues, in regimes with the required strengths of 
non-linearity \cite{Shapiro06} \cite{Shapiro07} \cite{Gea-Banacloche10}. 
\newline \indent Solid state qubits are also a very promising direction for the development of a quantum computer, with superconducting qubits coupled to quantum microwave modes forming particularly relevant examples \cite{Kirchmair12} \cite{Reed11}  \cite{Majer07}. There is potential for the generation of non-linearities in and between microwave modes, leading to cross-Kerr type, and other, interaction terms. However, in addition, when the qubit-field interaction is of the Jaynes-Cummings form and the dispersive limit is employed \cite{Blais04}, the interaction between a qubit and a field mode takes an equivalent form to the cross-Kerr non-linearity with
\begin{equation} \label{Hintss} H_{int}= \hbar \chi \sigma_{z_j} a^{\dagger} a  \end{equation}
where $a$ ($a^{\dagger}$) corresponds to the bus mode as above and $\sigma_{z_j}$ is the standard Pauli operator for a solid state qubit in mode $j$ \cite{Spiller07} \cite{Schoelkopf04} \cite{Vidiella-Barranco05} \cite{Milburn06}. During this paper we will use the solid state notation, taking the computational basis to be $\ket{0} \equiv \ket{\uparrow_z}$ and $\ket{1} \equiv \ket{\downarrow_z}$ where these are the eigenvectors of $\sigma_z \equiv \ket{\uparrow_z}\bra{\uparrow_z} - \ket{\downarrow_z}\bra{\downarrow_z}$. In direct analogy to the photonic qubit case, when this Hamiltonian (\ref{Hintss}) is applied for a time $t$, equal and opposite bus phase rotations of $\pm \theta$ are induced, dependent upon the state of the control atomic qubit.

We now present the desired criteria for an ideal two-qubit phase gate:
\begin{enumerate}
\item Maximally entangling. Two such gates are the CNOT and C-Phase gate.
\item No measurements. This implies that no operations dependent on the outcome of a measurement should be required either during implementation, or to herald the success, of the gate. This simplifies the procedure and may help to speed up the operation of the gate.
\item No inherent decoherence. By this we mean that in the case of a bus mediated gate, the final state of the system, after the operation of the gate, must be a product state between the bus mode and the computational qubits pair. Any remaining entanglement between the bus mode and the qubits will create decoherence even when implemented ideally.
\item Some level of robustness against parameter variations, so that the gate is still viable in the realistic case of imperfect implementation. Clearly the greater the level of robustness the better.
\end{enumerate}

We now consider some previous methods of implementing a bus mediated phase gate using cross-Kerr type nonlinearities.
Due to the values of $\chi$ that can be fabricated in physical systems, only schemes which permit small angles of rotation (\(<<\pi\) ) are of practical interest. Previous work has used this Hamiltonian within the physical restraint of small angles, in the context of both photonic and solid state qubits, to probabilistically or near-deterministically create entanglement and implement two-qubit gates by the use of measurement and classical feed-forward \cite{Beausoleil05} \cite{Spiller07} \cite{Munro04}. These methods rely on single qubit rotations dependent on the outcome of some measurement on the bus mode in order to implement the desired gate \cite{Munro05}. Due to this, these methods clearly do not meet criteria 2. If we consider using the standard displacement operator for the bus mode
\begin{equation} D(\alpha)=\exp \left(\alpha a^{\dagger} +\alpha^* a \right) \label{D}\end{equation}
 in conjugation with the cross-Kerr type interactions then a measurement-free phase gate can be implemented by performing four controlled rotations each followed by a displacement operation. The final state of the system is not however an exact product state between the bus mode and the computational qubits pair and so an inherent source of decoherence is introduced \cite{Milburn06}. This clearly fails criteria 3.
 The work here presents a new form of measurement-free phase gate which is a fundamentally improved method satisfying all the above criteria for any magnitude of rotation angles. We will also see that this gate is more robust to errors in implementation in a comparison to a previous method. First of all the new method will be outlined in its most general case using geometric arguments. From here the most simple specific example is studied and compared to the previous method.

\section{General case of the measurement-free phase gate between two qubits}
\label{perfect2qubit}
\begin{figure}[h!]
\center
\includegraphics[scale=0.4]{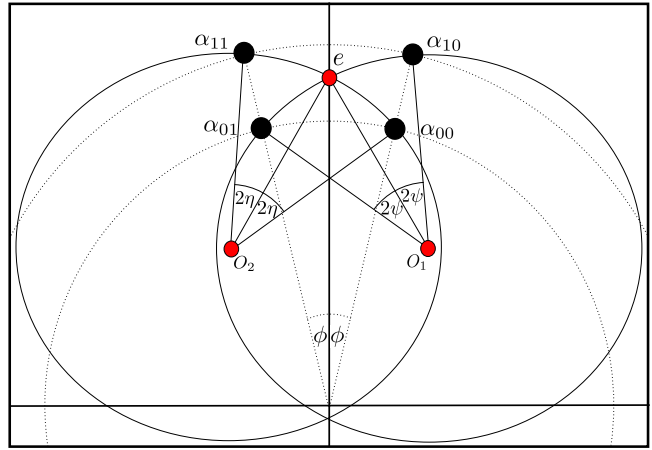}
\caption{\label{equidistances1} (Color online) The state after the 3rd operation. It is not possible to represent the most general displacement, described in the geometric method, graphically and so a simple example is chosen here that forces the point $e$ to be on the imaginary axis and $\eta=\psi$. This however is not the only possible geometry.}
\end{figure}

\begin{figure}[h!]
\center
\includegraphics[scale=0.4]{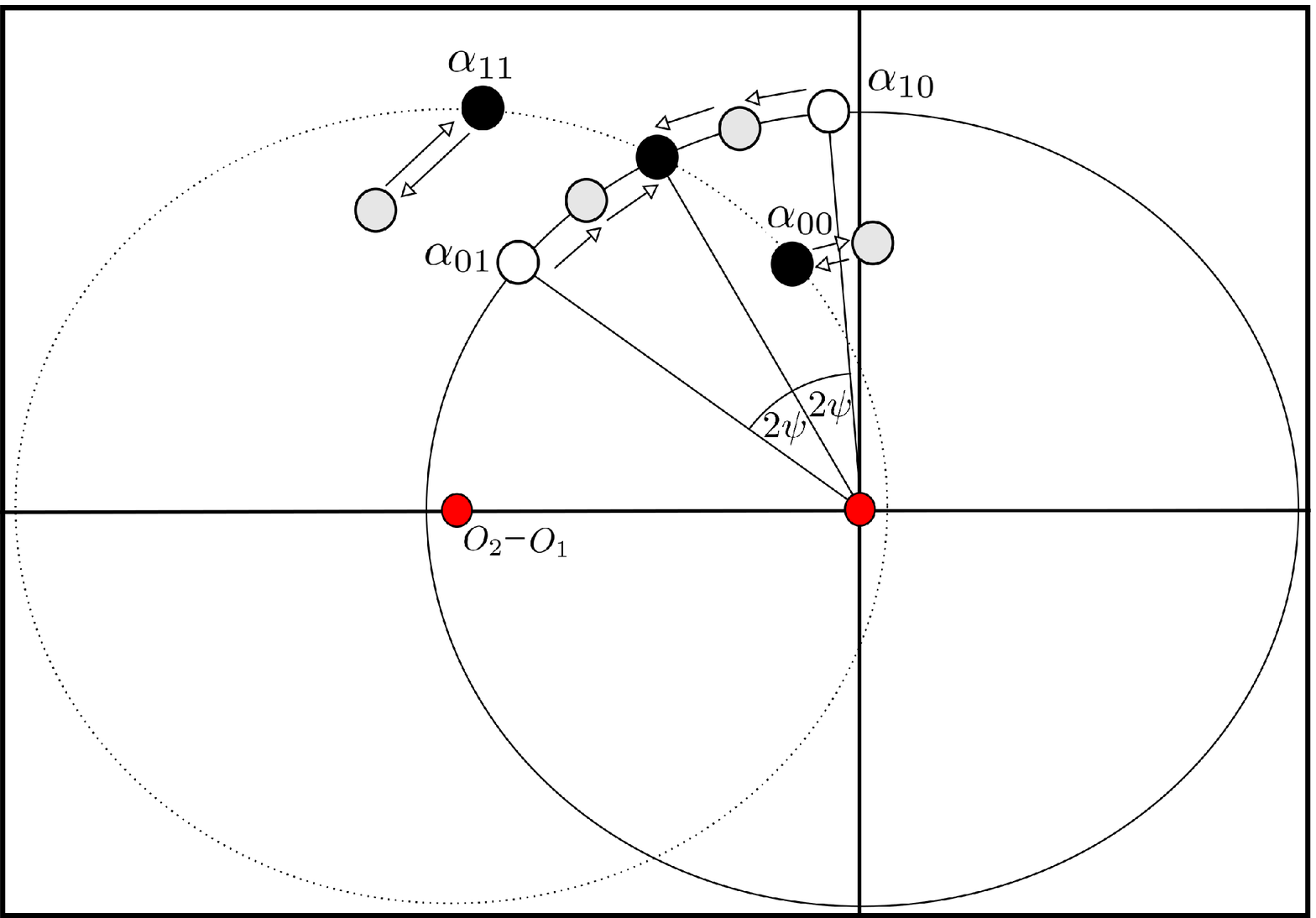}
\caption{\label{equidistances2} (Color online) The state before and after the 5th step. The white points show the state before the rotations (for $\alpha_{11}$ and $\alpha_{00}$ this is the same as the resultant state so these are not shown). The grey points show the state after the first rotation. The arrows show the effects of the rotations. The two rotations are of the form $e^{-i\psi\sigma_{z_1} a^{\dagger} a }$ and $e^{i\psi\sigma_{z_2} a^{\dagger} a }$.}
\end{figure}

\begin{figure}[h!]
\center
\includegraphics[scale=0.4]{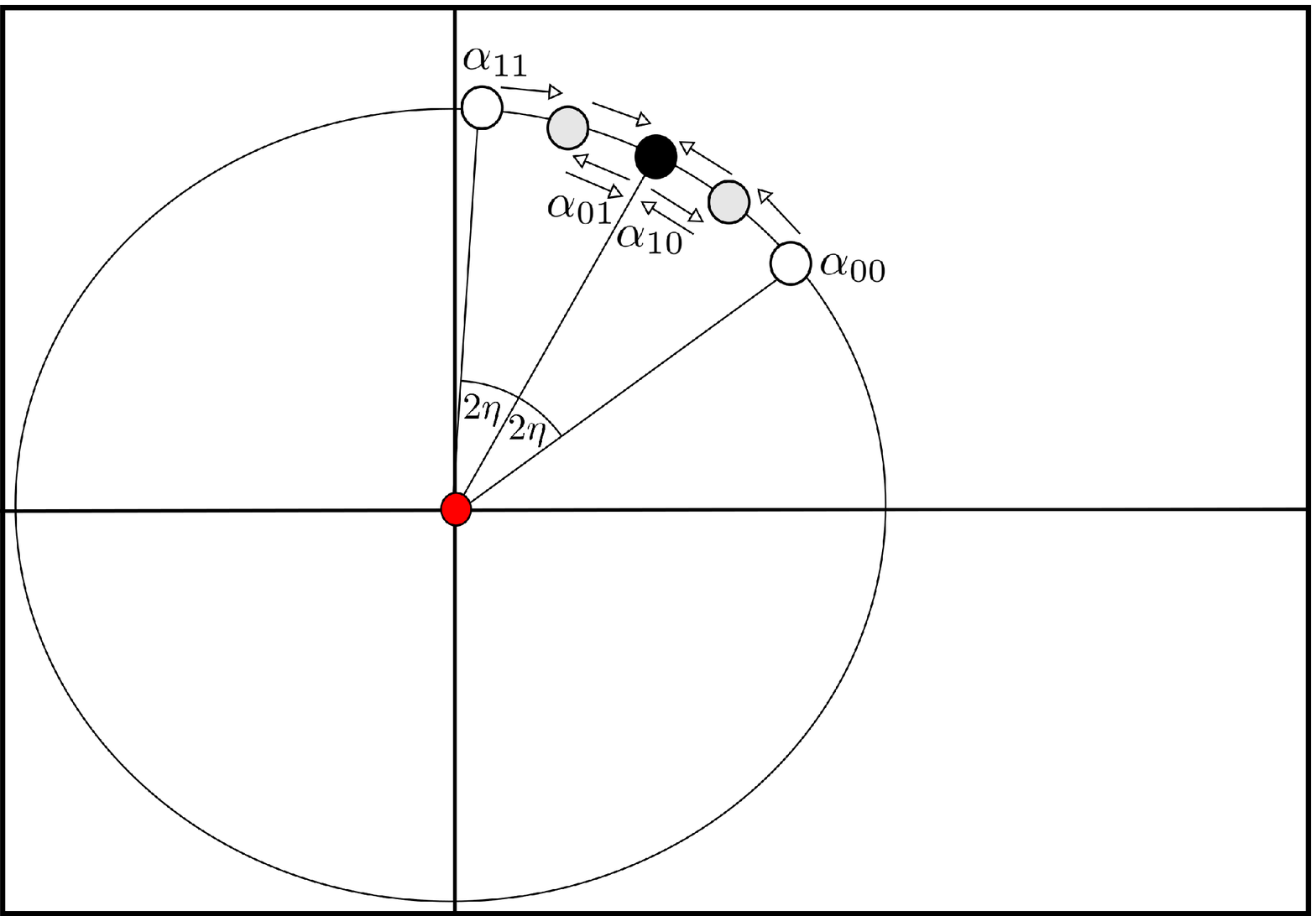}
\caption{\label{equidistances3} (Color online) The state before and after the 7th step.  The white points show the state before the rotations (for $\alpha_{01}$ and $\alpha_{10}$ this is the same as the resultant state so these are not shown). The grey points show the state after the first rotation. The arrows show the effects of the rotations. The two rotations are of the form $e^{i\eta\sigma_{z_1} a^{\dagger} a }$ and $e^{i \eta\sigma_{z_2} a^{\dagger} a }$.}
\end{figure}

Here we present a protocol whereby the bus mode is exactly disentangled after the implementation of the gate. We start by taking the most general input state, with the bus mode disentangled prior to any interactions. This is of the form
\begin{equation} \label{instate2q} \ket{\psi_0}=
\sum_{j,k=0,1}c_{jk}\ket{j_1k_2}\ket{\alpha}
\end{equation}
with the usual normalisation constraints on the coefficients $c_{jk}$ and the labels $1$ and $2$ denoting the two qubits (from now on the position in the ket will denote the qubit number).
 The initial state of the bus mode can be represented by a point in phase space at $\alpha$. A cross-Kerr rotation can only change the state of the bus mode (although this change is dependent on the state of the qubits). A displacement operation (\ref{D}) can only do this (independent of the state of the qubits) and in addition create a phase (dependent on the state of the bus mode). The state of the system after each operation can therefore be written in the form
\begin{equation}
\ket{\psi_t}=\sum_{j,k=0,1}c_{jk}e^{i\eta_{jk}}\ket{jk}\ket{\alpha_{jk}} \label{psit}
\end{equation}
which is in general entangled between bus and qubits.
The state of the bus mode can thus be represented by the four points in phase space $\alpha_{00}$, $\alpha_{01}$, $\alpha_{10}$ and $\alpha_{11}$. We therefore have the aim of implementing a controlled phase gate (which requires creating suitable $\eta_{jk}$ terms) whilst creating a final state where the bus is disentangled from the computational qubits, i.e.  $\alpha_{00}=\alpha_{01}=\alpha_{10}=\alpha_{11}$.
The method developed for achieving this can be most simply described geometrically in phase space and is outlined in the following 8 steps:
\begin{enumerate}
\item Apply a qubit 1 controlled rotation. This is an operator of the form $e^{-i\theta \sigma_{z_1} a^{\dagger} a }$.
\item Apply a displacement operator. The only restriction is that the argument of this operator is not a constant real multiple of $\alpha$. This is of the form $D(\omega)$ where $\omega \neq c\alpha$ for any $c$ where $c$ is real.
\item Apply a qubit 2 controlled rotation. This is an operator of the form $e^{-i\phi\sigma_{z_2} a^{\dagger}a}$. The resulting state after this operation is shown in figure \ref{equidistances1}. (There is also nothing that prohibits the additional implementation of a qubit 1 controlled rotation at this point).

\item Apply a displacement such that the point $O_1$ in figure \ref{equidistances1} is at the origin in phase space. This point is the origin of a circle on which $ \alpha_{01} $, $ \alpha_{10} $ and $e $ are situated. The point $ e $ is the unique point that is equidistant from $ \alpha_{01} $ and $ \alpha_{10} $ and also equidistant (although not necessarily the same distance as from $\alpha_{01}$ and $\alpha_{10}$) from $ \alpha_{00} $ and $ \alpha_{11} $. Due to this, after this displacement, the angles between $ \alpha_{01} $ and $ e $ and between $ e $ and $ \alpha_{10} $ (with respect to the origin) are the same and this is absolutely essential to the next step.

\item Apply a pair of rotations such that the overall effect is no net rotation on the points $\alpha_{00}$ and $\alpha_{11}$ and an equal and opposite rotation on the points $\alpha_{01}$ and $\alpha_{10}$ of the required magnitude such that they are rotated on to the point $ e'=e-O_1$. These controlled rotations are of the form $e^{-i\psi\sigma_{z_1} a^{\dagger} a }$ and $e^{i\psi\sigma_{z_2} a^{\dagger} a }$ where $\psi$ is half of the angle between $\alpha_{01}$ and $e$ (and that between $e$ and $\alpha_{10}$). After this pair of operations we have a state such that $\alpha_{01}=\alpha_{10}$ ($= e'$). These operations and the resultant state are shown in figure \ref{equidistances2}.

\item Apply a displacement such that the point $O_2-O_1$ in figure \ref{equidistances2} is at the origin in phase space. Similarly to step 4 this point is the origin of a circle on which $ \alpha_{00} $, $ \alpha_{11} $ and the point $\alpha_{01}=\alpha_{10} $ are situated. As before we have that both  $ \alpha_{00} $ and  $ \alpha_{11} $ are equidistant from the point $\alpha_{01}=\alpha_{10} $ and so the angle between  $ \alpha_{11} $ and  $ \alpha_{01} = \alpha_{10} $ is equal to that between  $ \alpha_{01} = \alpha_{10} $ and  $ \alpha_{00} $. The equality of these angles is essential to the next step. 

\item Apply a pair of rotations such that the overall effect is no net rotation on the points $\alpha_{01}$ and $\alpha_{10}$ and an equal and opposite rotation on the points $\alpha_{00}$ and $\alpha_{11}$ of the required magnitude such that they are rotated on to the point $ \alpha_{01}=\alpha_{10} $. These controlled rotations are of the form $e^{i\eta\sigma_{z_1} a^{\dagger} a }$ and $e^{i \eta\sigma_{z_2} a^{\dagger} a }$ where $\eta$ is half of the angle between $\alpha_{11}$ and $ \alpha_{01}=\alpha_{10} $ (and that between $ \alpha_{01}=\alpha_{10} $ and $\alpha_{00}$). After this pair of operations we have a state such that the $\alpha_{00}=\alpha_{01}=\alpha_{10}=\alpha_{11}$, i.e. the bus mode is disentangled. These operations and the resultant state are shown in figure \ref{equidistances3}.

\item This is a final optional displacement so that the bus mode is back at its initial point in phase space. This may be necessary if performing multiple gates in succession, however the qubits are already disentangled so it is not essential for the implementation of the gate.

\end{enumerate}

\begin{figure*}
\includegraphics[scale=0.5]{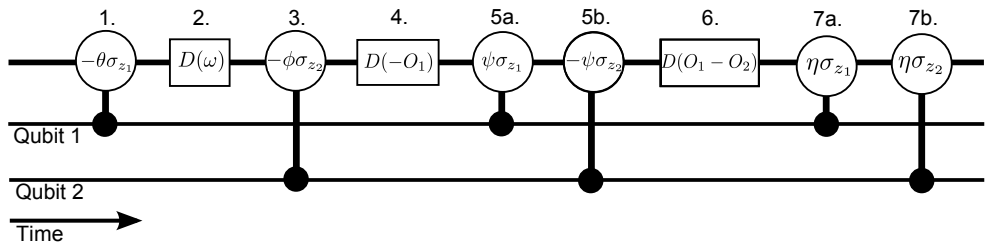}
\caption{\label{circuitdiagram}A circuit diagram of the general method described herein for implementing a measurement-free phase gate. $\theta$, $\phi$, $\psi$ and $\eta$ take real values and $\omega$, $O_1$ and $O_2$ take complex values. Given $\theta$, $\phi$ and $\omega$, the remaining coefficients are chosen to obey this general method, i.e. they are functions of $\theta$, $\phi$ and $\omega$.}
\end{figure*}

These steps are summarised in the circuit diagram of figure~\ref{circuitdiagram}.

\section{\label{specificexample}Specific example of the measurement-free phase gate between two qubits}
It is now necessary to show that the operations outlined above can be chosen such that the phases created perform the desired C-Phase gate. The other question yet to be addressed is whether this method will allow a feasible choice of operations; we have not shown that choosing the first two rotations to be small does not force at least one of the remaining displacements or rotations to be impractically large. To show that the required phases can be created, and that this can be done with a feasible set of operations, it is simpler to deal with a specific set of operators that can be derived from this geometric recipe. Before doing this it is worth noting that the whole method is defined by the choices made in the first 3 steps. Due to this, it is straightforward to consider a very (although not the most) general case. Here we consider the case where the displacement in step 2 is of a form such that the point $e$ is on the imaginary axis (given that $\alpha$ is real) - see figure \ref{equidistances1}. Taking this as our choice of displacement it is possible to derive, using basic geometry, that the set of operations that obey the above method are:
\begin{enumerate} 
\item  $ e^{-i\theta a^{\dagger}a \sigma_{z_1}}$. 
\item $D\left(\beta i - \alpha \cos\theta\right) $. This is taking $\omega=\beta i - \alpha \cos\theta$.
\item  $e^{ -i\phi\sigma_{z_2} a^{\dagger} a }$. 
\item \( D\left(- \frac{\alpha \sin\theta}{2 \sin\phi} -\frac{\beta}{2 \cos\phi}i\right)\).
\item  $ e^{-i\phi\sigma_{z_1} a^{\dagger} a }$ and  $ e^{i\phi\sigma_{z_2} a^{\dagger} a }$.
\item  \(D\left(\frac{\alpha \sin\theta}{\sin\phi}\right) \).
\item  $ e^{i \phi\sigma_{z_1} a^{\dagger} a }$ and $ e^{i \phi\sigma_{z_2} a^{\dagger} a }$. 
\item  \(D\left(\alpha-\frac{\alpha \sin\theta}{2\sin\phi}-\frac{\beta}{2 \cos\phi}i\right) \)  (Optional).
\end{enumerate}
where $\alpha$, $\beta$, $\theta$ and $\phi$ are real and $\alpha$ is the value defining the initial state of the bus mode.
\newline \indent From this set of operations it is straightforward to confirm that the final state will leave the bus mode disentangled and in its initial state. This is clearly necessary for the method to be of any use, however we also require that a C-Phase gate has been implemented. In order to enforce the required condition for this we calculate the concurrence of the final state of the two qubits given that the initial state is the equal superposition state, that is $c_{jk}=\frac{1}{2}$ for each $j$ and $k$ in (\ref{instate2q}). The concurrence of a bi-partite pure state written in the form of (\ref{psit}), but with the bus disentangled, is given by
\begin{equation} C(\psi_t) = 2 \left|  c_{00}c_{11} e^{i (\eta_{00} + \eta_{11})}- c_{01}c_{10} e^{i (\eta_{01} + \eta_{10})} \right|.\end{equation}
This takes the value zero when there is no bi-partite entanglement and unity when there is maximum bi-partite entanglement \cite{Hill97}.
A detailed calculation gives
\begin{equation}
\label{conequi}
C^2= \frac{1}{2} - \frac{1}{2}\cos(4\alpha\beta\sin\theta\tan\phi).
\end{equation}
If a C-Phase gate has been implemented on this initial state we will end up in a maximally entangled state. Taking the solution of (\ref{conequi}) $=1$ we clearly get maximum entanglement when
\begin{equation}
\label{maxcon}
\alpha\beta\sin\theta\tan\phi=(2n+1)\frac{\pi}{4} \hspace{0.5cm}
\end{equation}
where $n$ is an integer. This protocol, combined with some local rotation on each qubit, therefore performs a C-Phase gate. 
\newline \indent In contrast to the previous methods that do not require measurement (see reference \cite{Milburn06}), if this is ideally implemented, criteria 3 is satisfied, i.e. the bus is exactly disentangled and so there is no inherent decoherence due to tracing out the bus mode. However it is important to check the concurrence for robustness to errors in the parameters $\alpha$, $\beta$, $\theta$ and $\phi$.
A Taylor expansion in each of the 4 variables around the points of maximum concurrence (taking $n=0$) gives the equations
\begin{equation}   C^2\simeq 1 -\frac{\pi^2}{4}\left(\frac{\alpha_{E}}{\alpha_M}\right)^2  \label{alpha'} \end{equation}
\begin{equation}   C^2\simeq 1 -\frac{\pi^2}{4}\left(\frac{\beta_E}{\beta_M}\right)^2 \label{beta'} \end{equation}
\begin{equation} C^2\simeq 1 - \frac{\pi^2}{4}\left(\frac{\theta_{E}}{\tan\theta_M}\right)^2   \label{theta'} \end{equation}
\begin{equation}  C^2\simeq 1 - \frac{\pi^2}{2}\left(\frac{ \phi_E}{\sin(2\phi_M)}\right)^2  \label{phi'} \end{equation}
where $\alpha_E = \alpha - \alpha_M$ with $\alpha_M$ obeying the maximum concurrence equation (\ref{maxcon}) (and similarly for the $\beta$, $\theta$ and $\phi$). Considering (\ref{alpha'}) and (\ref{beta'}) we see that to obtain $C > 0.97$ a relative error of up to $\approx$ 0.155 can be tolerated in both $\alpha$ and $\beta$. Conversely a 0.01 relative error in either of these will result in $C \approx 0.99987$. The concurrence is therefore clearly very robust to errors in the parameters associated with the displacements. We now consider equations (\ref{theta'}) and (\ref{phi'}). If we take the specific solution to (\ref{maxcon}):
\[ \alpha=\beta=11.0719... \hspace{0.5cm} \theta=\phi=0.08\]
then we get the following equations
\[ C^2=1-194\phi_{E}^2 \hspace{1cm} C^2=1-383\theta_{E}^2.\]
With these values, to obtain $C>0.97$ we can tolerate an error of up to $\approx$ 0.012 in $\theta$ (a 0.15 relative error). Conversely an error of 0.001 in $\theta$ (a 0.0125 relative error) will result in $C \approx 0.9998$. The concurrence is less sensitive to errors in $\phi$. Here we have chosen small values for $\alpha$ and $\beta$ but these are still experimentally feasible - the mean number of photons in a coherent state $\ket{\alpha}$ is $|\alpha|^2$ which in this case is $\approx$ 120 photons \cite{Knight}. There is a degree of flexibility over which variables have the greater error tolerance. An optimum choice of values for an implementation would depend on the system on which this was to be implemented and any constraints that this system might impose. Given this information, a system specific optimum could be found. From this example the link to conditional displacements can be seen, as in \cite{Wang02}. 

\section{\label{squareerrors} Comparison to a previous method}
We shall first describe in more detail a previous method of implementing a C-Phase gate without measurement that was introduced earlier \cite{Milburn06}. It is implemented by performing four controlled rotations each followed by a displacement operation. All the rotations are chosen to be of equal magnitude and the first and third rotation are controlled by the first qubit and the second and fourth rotation by the second qubit. The displacements are chosen such that in the case of no rotations the bus mode would travel around a square (centered on the origin) in phase space and so end up back in its initial state.  The rotations cause the bus mode associated with the four two-qubit basis states to travel around slightly different paths in phase space and it is the difference between these closed path areas that create the geometric phases that can be chosen appropriately to implement a C-Phase gate. If the rotations are taken to be small, they may be approximated by conditional displacements. In this approximation, after implementation of the gate, the final state of the system is a product state between the bus mode and the qubits, and so no measurement is required. However due to this approximation the bus is not precisely disentangled from the qubits and so tracing out the bus mode creates a small amount of decoherence even in the ideal case. Taking the expressions for the phases created by these operations calculated in \cite{Milburn06} and taking the small angle limit (which is necessary for the method to be valid) we can derive that the concurrence is given by
\begin{equation}C^2 \simeq \frac{1}{2}-\frac{1}{2}\cos(12\alpha^2\theta^2) \label{squarecon}\end{equation}
where $\alpha e^{i\frac{\pi}{4}}$ is the complex number characterising the initial coherent state, with $\alpha$ real, and $\theta$ is the angle of each rotation. This is a maximum at
\begin{equation}\alpha^2\theta^2=\frac{\pi}{12}(2n+1)\label{squaremax}\end{equation}
where $n$ is an integer. If we take Taylor expansions in each variable around the points of maximum entanglement, taking $n=0$, we get
\begin{equation} C^2 \simeq 1 - \pi^2\left(\frac{\alpha_E}{\alpha_M}\right)^2 \hspace{0.5cm} \label{alphasquare'} \end{equation}
\begin{equation} C^2 \simeq 1 - \pi^2\left(\frac{\theta_E}{\theta_M}\right)^2 \hspace{0.5cm} \label{thetasquare'} \end{equation}
where  \(\alpha_{M}$ and $\theta_{M}\) obey equation (\ref{squaremax}) with $\alpha_{E}=\alpha-\alpha_{M}$ and $\theta_E=\theta-\theta_M$. Considering both (\ref{alphasquare'}) and (\ref{thetasquare'}), we see that to obtain $C > 0.97$ we can tolerate a relative error in either $\alpha$ or $\theta$ of up to $\approx 0.077$. The new method described here can tolerate twice this error in the equivalent parameters.
\newline \indent We can furthermore consider the inherent bus error that this method would incur if the specific values of $\alpha$ and $\theta$ taken in section~\ref{specificexample} are used. It can be shown that maximum magnitude of the difference in state of the bus mode associated with the different two qubit states is $|\alpha_{00_{f}}-\alpha_{01_{f}}|\approx 0.2$ where $\alpha_{ij_{f}}$ is the final bus state associated with the two qubit state $\ket{ij}$ ( $|\alpha_{00_{f}}-\alpha_{11_{f}}|= 0 $ and $|\alpha_{01_{f}}-\alpha_{10_{f}}|< 0.2 $ ). This can alternatively be expressed in terms of the initial magnitude of the bus state by $|\alpha_{00_{f}}-\alpha_{01_{f}}|\approx 0.02|\alpha|$, which is a difference of about $2\%$. This will therefore introduce decoherence effects, even in the ideal case. This is an additional error that the new method does not suffer from.
\newline

\section{Conclusions}

We have discussed the desired criteria for a two-qubit phase gate and presented a general protocol that satisfies these in the context of qubus computation using a cross-Kerr type non-linearity between the bus and the computational qubits. We have shown, using a specific example, that this method is robust against parameter variations and therefore is low error. This is relevant not only to optical qubits but also to solid state qubits interacting with the bus via the Jaynes-Cummings interaction in the dispersive limit. This method is extremely general as it is not only adaptable to different hardware but because the operations themselves can be chosen to optimise the gates performance dependent upon the constraints imposed by the physical system. Furthermore this method fundamentally improves on previous similar methods due to the exact geometry employed which results in no decoherence errors in the ideal limit. 
\newline \indent We have presented a general and realisation-independent approach in this paper, to illustrate its wide applicability. With some chosen specific realisation, possible extensions of this work could be to model the effects of loss on the bus mode \cite{Louis08} and system specific decoherence on the qubits. It may also be interesting to then investigate the optimum choice of operations for a particular realisation.


\begin{thebibliography}{99}

\bibitem{Milburn06} T. P. Spiller, K. Nemoto, S. L. Braunstein, W. J. Munro, P. van Loock and G. J. Milburn, New J. Phys. {\bf  8}, 30 (2006)

\bibitem{Braunstein05} S.L. Braunstein and P. van Loock, Reviews of Modern Physics {\bf 77}, 513 (2005).

\bibitem{Lloyd99} S. Lloyd and S. L. Braunstein, Phys. Rev. Lett. {\bf 82}, 1784 (1999).

\bibitem{Ekert95} D. Deutsch, A. Barenco and A. Ekert, \emph{Proceedings of the Royal Society London A} {\bf 449}, 669 (1995).

\bibitem{Gerry01} C. C. Gerry and R. A. Campos,  Phys. Rev. A {\bf 64}, 063814 (2001).

\bibitem{d'Ariano00} G. M. D'Ariano, C. Macchiavello and L. Maccone,  Fortschr. Phys. {\bf 48}, 573 (2000).

\bibitem{Kok02} P. Kok, H. Lee and J. P. Dowling, Phys. Rev. A {\bf 66}, 063814 (2002)

\bibitem{Yang09} X. Yang, S. Li, C. Zhang, and H. Wang, J. Opt. Soc. Am. B {\bf 26}, 1423 (2009).

\bibitem{Sun08} H. Sun, Y. Niu, S. Jin and S. Gong, J. Phys. B: At. Mol. Opt. Phys. {\bf 41}, 065504 (2008).

\bibitem{Spiller05} W. J. Munro, K. Nemoto, R. G. Beausoleil and T. P. Spiller, Phys. Rev. A {\bf 71}, 033819 (2005).

\bibitem{Harris99} S. E Harris, and L. V. Hau, Phys. Rev. Lett. {\bf 82}, 4611 (1999).

\bibitem{Matsuda09} N. Matsuda, R. Shimizu, Y. Mitsumori, H. Kosaka
and K. Edamatsu, Nature Photonics {\bf 3}, 95 (2009).

\bibitem{Li04} X. Li, P. L. Voss, J. E. Sharping and P. Kumar, Phy. Rev. Lett. {\bf 94}, 053601 (2005)

\bibitem{Zhu10} Y. Zhu, Optics Lett. {\bf 35}, 303 (2010).

\bibitem{Mucke10} M. M\"{u}cke, E. Figueroa, J. Bochmann, C. Hahn, K. Murr, S. Ritter,
C. J. Villas-Boas and G. Rempe, Nature, {\bf 465}, 755 (2010).

\bibitem{Dao-Ming07} L. Dao-Ming and Z. Shi-Biao, Chinese Phys. Lett. {\bf 24}, 1567 (2007).

\bibitem{Shapiro06} J. H. Shapiro, Phys. Rev. A {\bf 73}, 062305 (2006). 

\bibitem{Shapiro07} J. H. Shapiro and M. Razavi, New J. Phys. {\bf 9}, 16 (2007).

\bibitem{Gea-Banacloche10} J. Gea-Banacloche, Phys. Rev. A {\bf 81}, 043823 (2010). 

\bibitem{Kirchmair12} Kirchmair \emph{et al.} arxiv:1211.2228 (2012).

\bibitem{Reed11} M. D. Reed, L. DiCarlo, S. E. Nigg, L. Sun, L. Frunzio, S. M. Girvin and R. J. Schoelkopf, Nature {\bf 482}, 382 (2012).

\bibitem{Majer07} J. Majer \emph{et al.} Nature {\bf 449}, 443 (2007).

\bibitem{Blais04} A. Blais, R. S. Huang, A. Wallraff, S.M. Girvin and R.J. Schoelkopf, Phys. Rev. A {\bf69}, 062320 (2004).

\bibitem{Spiller07}S.G.R. Louis, K. Nemoto, W.J. Munro and T.P. Spiller, New J. Phys. {\bf 9}, 193 (2007).

\bibitem{Schoelkopf04} A. Blais, R. S. Huang, A. Wallraff, S. M. Girvin and R. J. Schoelkopf, Phys. Rev. A {\bf 69}, 062320 (2004).

\bibitem{Vidiella-Barranco05} F. L. Semi\~{a}o and A. Vidiella-Barranco, Phys. Rev. A {\bf 72}, 064305 (2005).

\bibitem{Beausoleil05}W.J. Munro, K. Nemoto, T.P. Spiller, S.D. Barrett, P. Kok and R.G. Beausoleil, J. Opt. B: Quantum Semiclass. Opt. {\bf 7} S135 (2005).

\bibitem{Munro04} K. Nemoto and W. J. Munro, Phys. Rev. Lett. {\bf 93}, 250502 (2004).

\bibitem{Munro05} K. Nemoto and W. J. Munro, Phys. Lett. A {\bf 344}, 104 (2005).

\bibitem{Hill97} S. Hill and W. K. Wootters, Phys. Rev. Lett. {\bf 78}, 5022 (1997).

\bibitem{Knight} C. C. Gerry and P. L. Knight, \emph{Introductory Quantum Optics}, (Cambridge
University Press, 2005), Ch. 3

\bibitem{Wang02} X. Wang and P. Zanardi, Phys. Rev. A {\bf 65}, 032327 (2002).

\bibitem{Louis08}  S.G. R. Louis, W. J. Munro, T. P. Spiller, and K. Nemoto, Phys. Rev. A {\bf 78}, 022326 (2008).


\end{thebibliography}
\end{document}